\documentclass[aps,prl,twocolumn,float]{revtex4}
\usepackage{amsmath,bm,epsfig}
\let\*\cdot
\def\<{\left\langle} \def\>{\right\rangle} \def\({\left(} \def\){\right)}
\let\p\partial \let\~\widetilde \let\^\widehat 

\def\be{\begin{equation}}\def\ee{\end{equation}}
\def\bea{\begin{eqnarray}}\def\eea{\end{eqnarray}}
\def\bse{\begin{subequations}}\def\ese{\end{subequations}}
\newcommand{\BE}[1]{\begin{equation}\label{#1}}
\newcommand{\BEA}[1]{\begin{eqnarray}\label{#1}}
\newcommand{\BSE}[1]{\begin{subequations}\label{#1}}

\usepackage{pstricks}
\usepackage{pst-node}
\usepackage[ansinew]{inputenc}
\usepackage{amssymb,amsmath}

\def\BSE{\begin{subequations}}\def\ESE{\end{subequations}}
\let \= \equiv

\def\p{\partial}

\def\e{\varepsilon}

\def\o{\omega}
\def\wt{\widetilde}

\def\be{\begin{equation}}       \def\ba{\begin{array}}

\def\ee{\end{equation}}         \def\ea{\end{array}}

\def\bea {\begin{eqnarray}}      \def\eea {\end{eqnarray}}

\def\bean{\begin{eqnarray*}}    \def\eean{\end{eqnarray*}}

\def\const {\mathop{\rm const}\nolimits}

\def\RA {\ \Rightarrow\ }

\def\<{\langle} \def\({\left(}  \def\>{\rangle} \def\){\right)}

\newtheorem{exi}{Example}

\begin{document}

\title{Discrete Wave Turbulence}
\author{Elena Kartashova}
 \email{Lena@risc.uni-linz.ac.at}
  \affiliation{RISC, J. Kepler University, Linz 4040, Austria}%%

   \begin{abstract}
In this Letter we present discrete wave turbulence (DWT) as a counterpart of classical statistical wave turbulence (SWT). DWT is characterized by resonance clustering, not by the size of clusters, i.e. it includes, but is not reduced to, the study of low-dimensional systems. Clusters with integrable and chaotic dynamics co-exist in different sub-spaces of the $\mathbf{k}$-space. NR-diagrams are introduced, a handy graphical presentation of an arbitrary resonance cluster allowing to reconstruct uniquely dynamical system describing the cluster.  DWT is shown to be a novel research field in nonlinear science, with its own methods, achievements and application areas.
\end{abstract}


%05.45.-a nonlinear dynamics
%Turbulent flows, 47.27.-i
%47.10.-g 	General theory in fluid dynamics
\maketitle

\noindent {\bf 1. Introduction}.
In 1960-1980th a great volume of research has been performed in the area of wave interactions \cite{waves}, comprehensive account of theory and experiment can be found in \cite{Cr85}. The notion of wave kinetic equation was introduced \cite{KinEq}; using some statistical assumptions, general methods
for deriving kinetic equations and their stationary solutions (energy spectra) were developed and statistical wave turbulence (SWT) theory was founded \cite{ZLF92}, with finite-size effects left aside. Their preliminary studies were performed in \cite{K94,AMS98}, where it was established that  nonlinear resonances are divided into dynamically independent, non-intersecting clusters. Explicit constructing of resonance clustering became the immediate challenge of great intricacy because no analytical methods for solving resonance conditions were known (the problem in its general form is equivalent to Hilbert's 10th problem \cite{Mat70}); brute-force computer computations do not help either, while integers to be dealt with, are too big. The problem has been recently solved \cite{KK06}. For complete resonance set, classical resonance curves \cite{curves} are not anymore a suitable representation, of course. Instead,  representation by a hyper-graph  on a plane was introduced  in \cite{KM07} (for 3-wave systems), which allows to extract uniquely the dynamical system describing each cluster. The study of resonant clustering yielded a model of laminated turbulence \cite{K06-1} which gives kinematic explanation of co-existence of STW and DWT in wave turbulent systems.  These and other results, both for 3- and 4-wave systems, are reviewed below, as well as their physical relevance and application areas.

\begin{figure}[h]
\includegraphics[width=6.5cm,height=3.5cm]{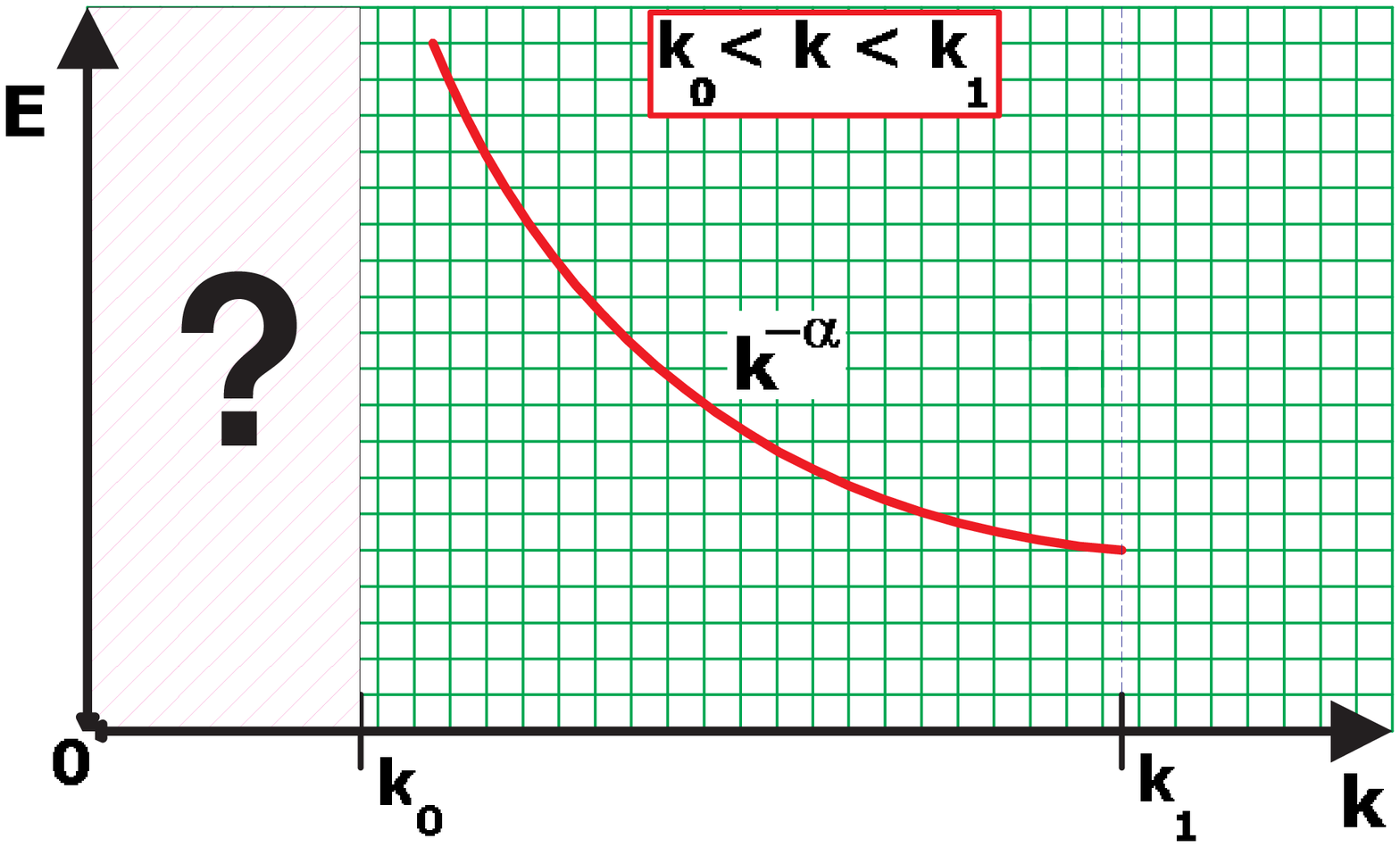}
\includegraphics[width=6.5cm,height=3.5cm]{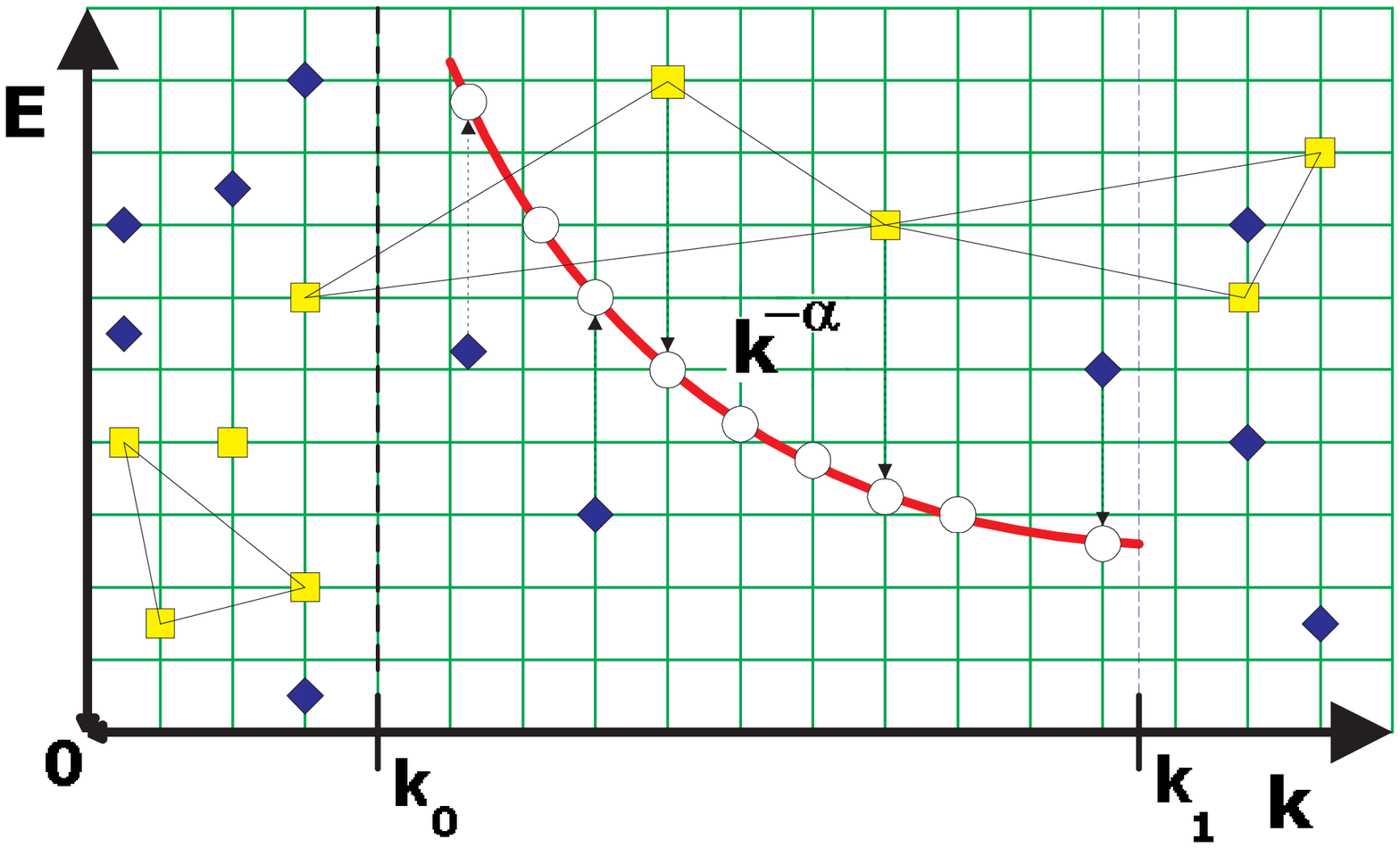}
\vskip -.4cm
\caption{\label{f:LamTur} Schematic representation of classical SWT (upper panel) and both SWT and DWT (lower panel). Here $E$, $k$ and $(k_0,k_1)$ are notations for energy, wave-number and  inertial interval correspondingly.}
\end{figure}

\noindent {\bf 2. Why  are predictions of SWT theory often not corroborated?}
The SWT theory
assumes weak nonlinearity, randomness of phases and infinite-box limit, i.e.
 the
resonance broadening $\Omega$ is greater than the spacing
$\delta_\omega$ between adjacent wave modes
\vskip -0.6cm \be
 \Omega > (\p \o
/\p k) 2 \pi/ L,
\label{quasi-r}
\ee
\vskip -0.3cm
where $L$ is the box size. Additionally, existence of inertial interval
$(k_o, k_1)$ is assumed, where energy input and dissipation are balanced.
If $k\gg k_1,$ dissipation suppresses nonlinear dynamics (see  Fig.\ref{f:LamTur}, the upper panel).  If $k\ll k_0$, finite-size effects take place which are due to boundary conditions and should be regarded separately.
\vskip -.4cm
\begin{figure*}
\begin{center}
\includegraphics[width=6.8cm,height=3cm]{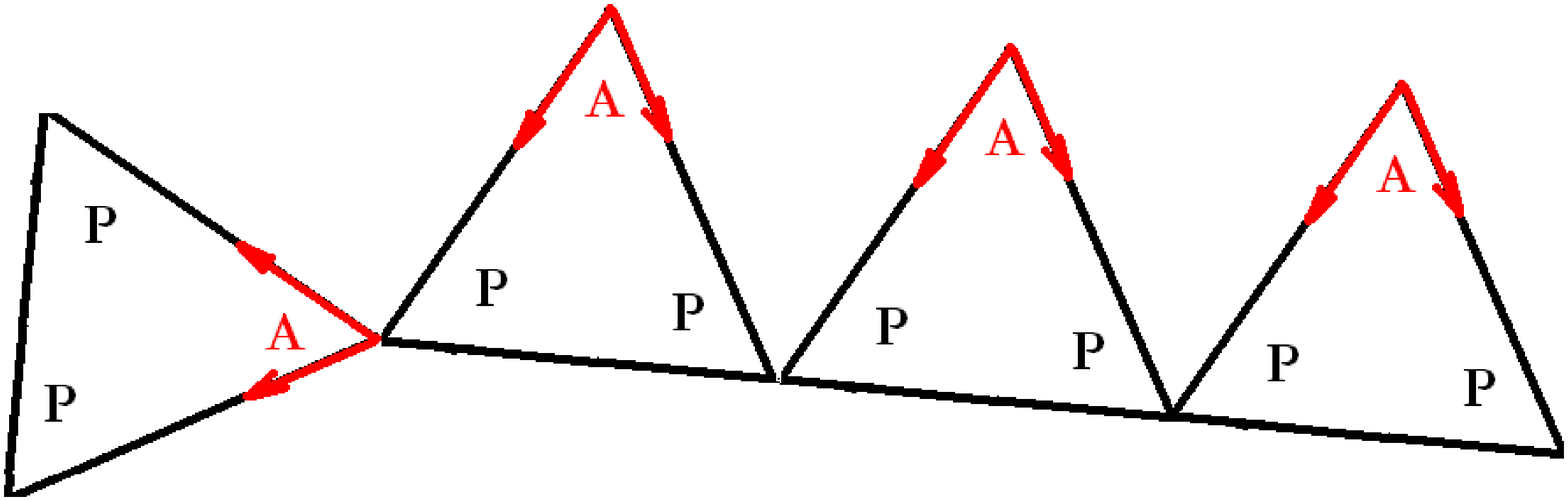}
\includegraphics[width=4.5cm,height=3cm]{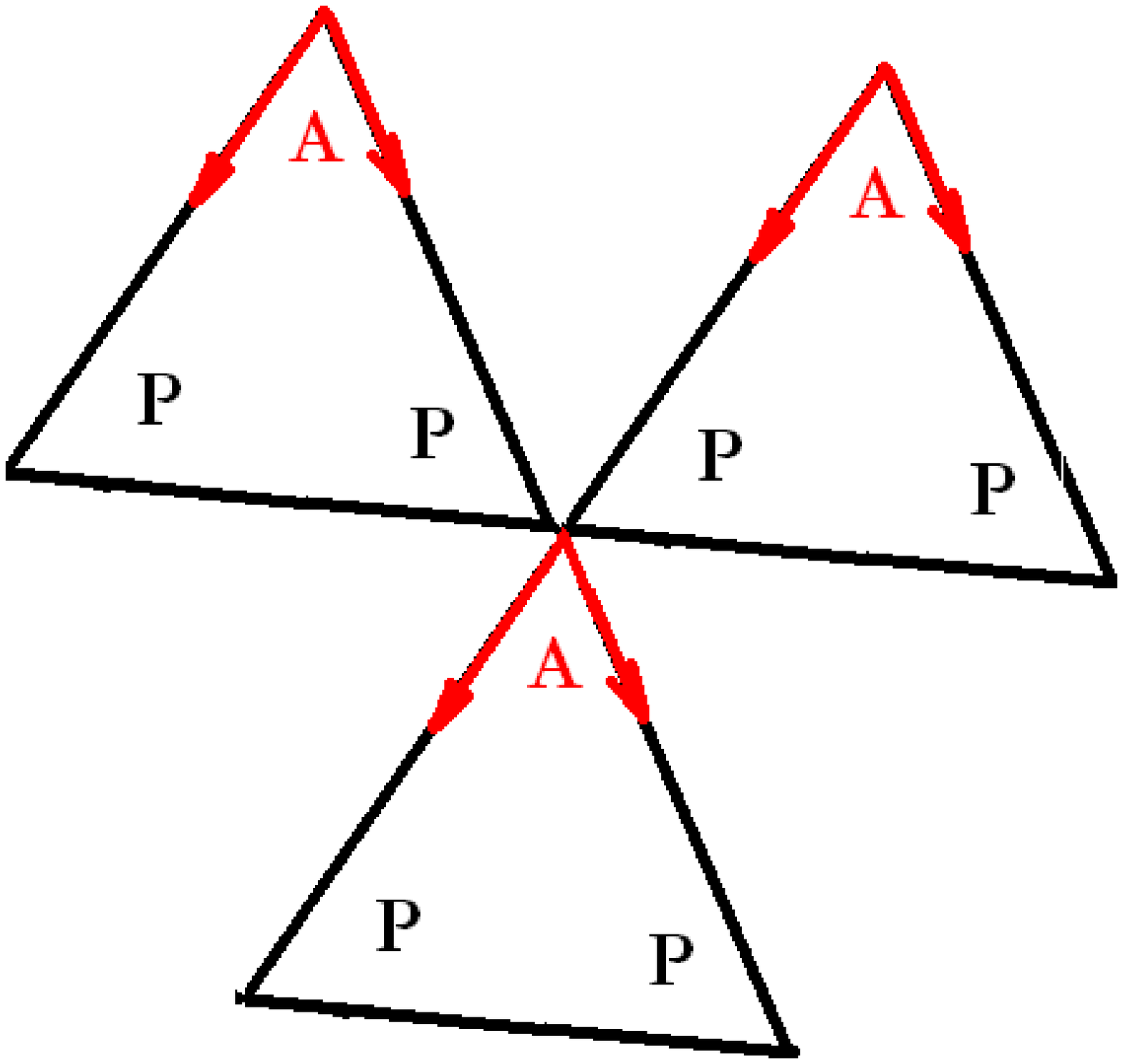}
\includegraphics[width=4cm,height=3cm]{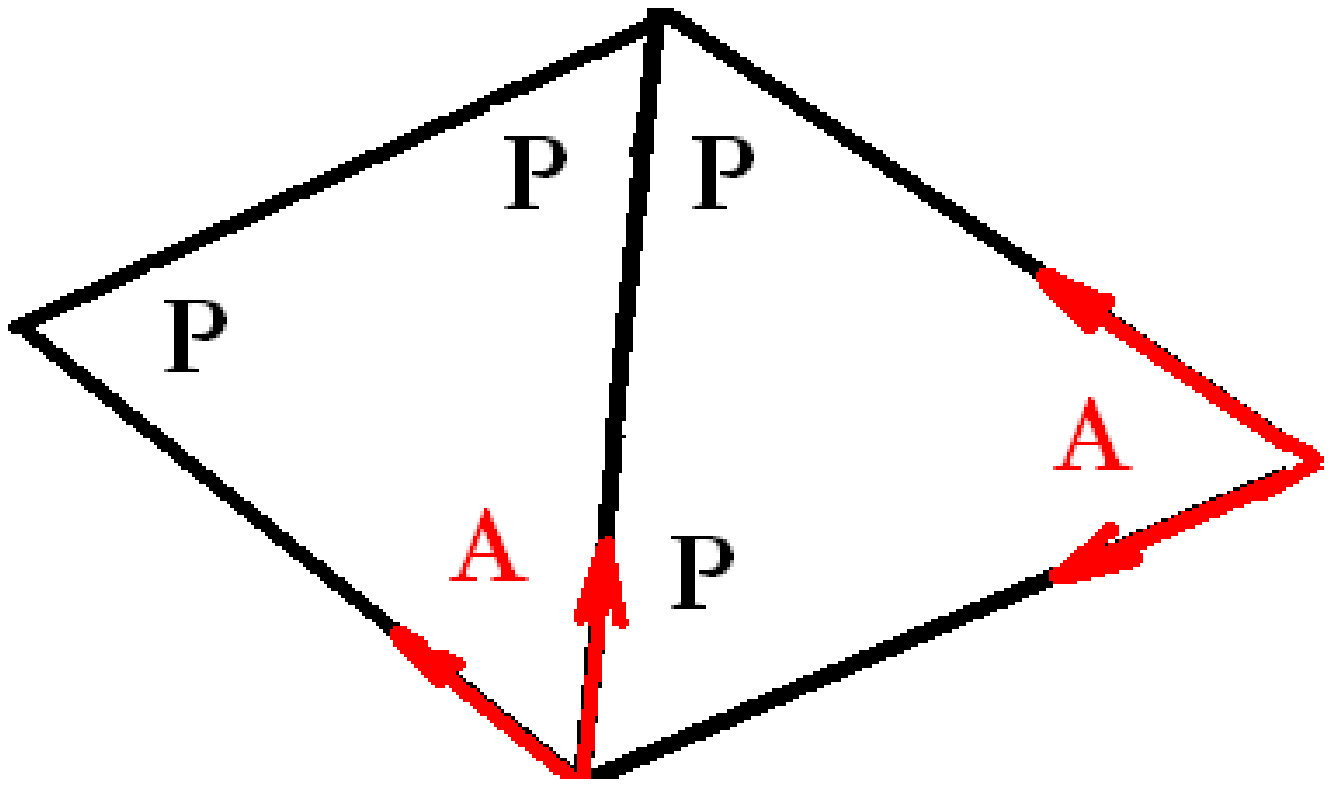}
\includegraphics[width=5.0cm,height=0.4cm]{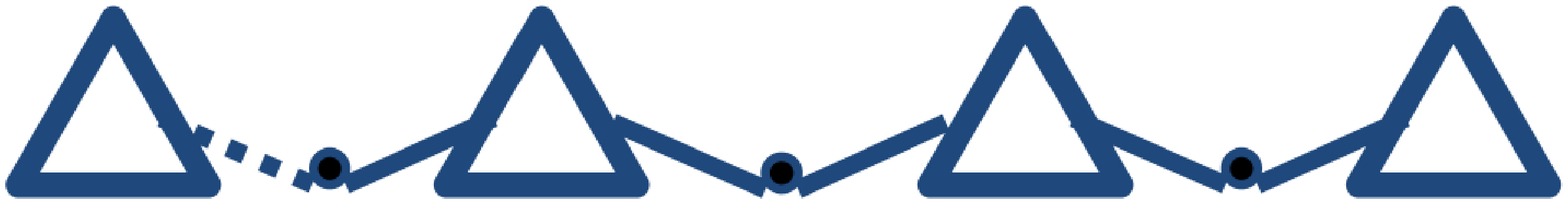}
\includegraphics[width=3cm,height=1.2cm]{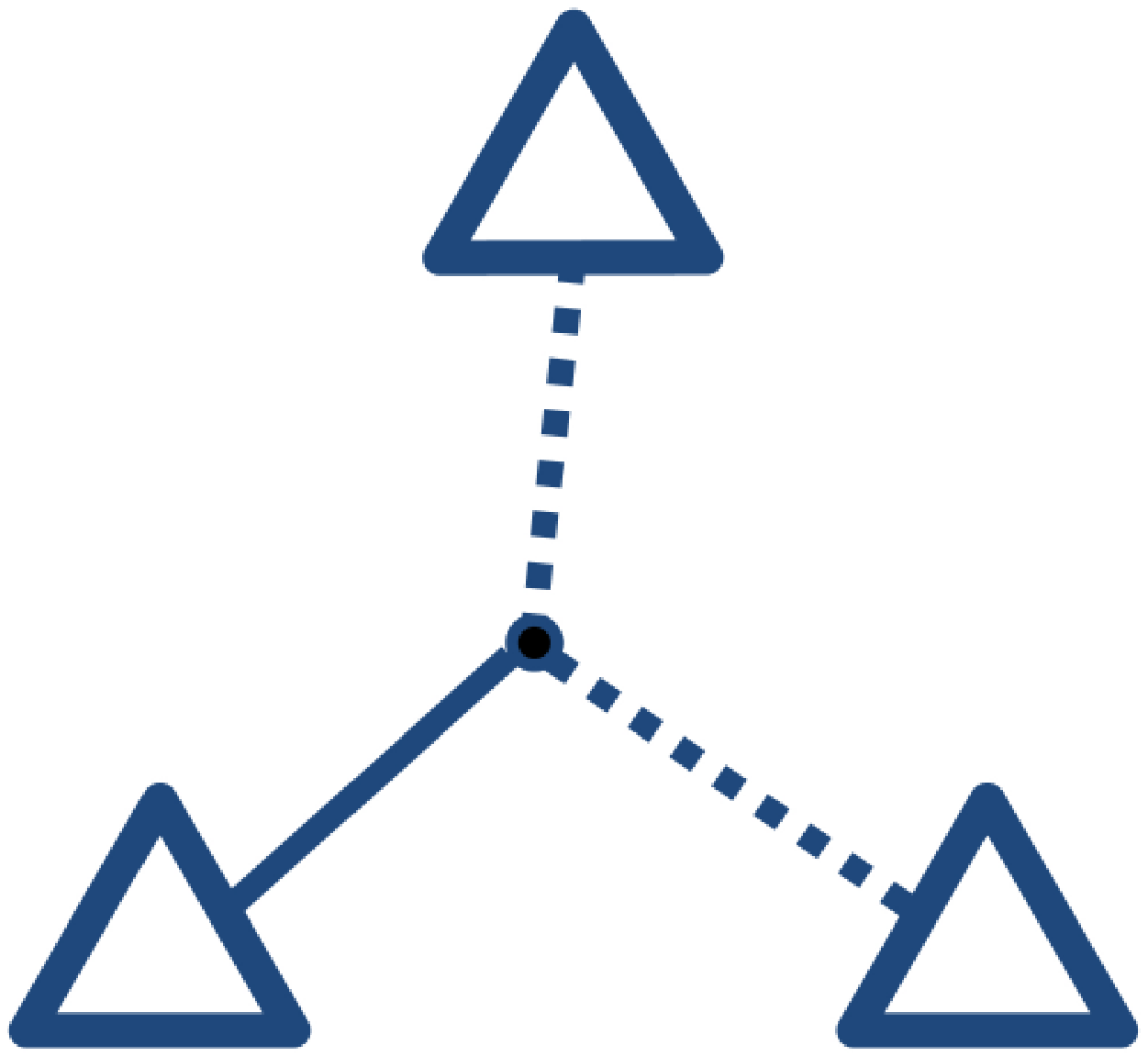}
\includegraphics[width=2.5cm,height=0.4cm]{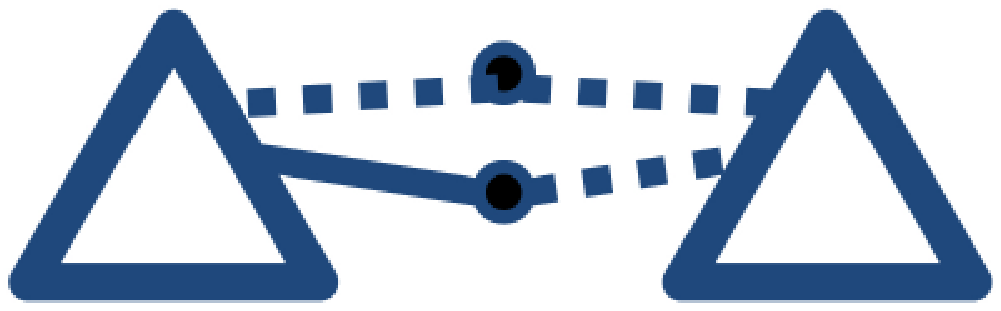}
\end{center}
\caption{\label{f:3Graph} Examples of clusters in a 3-wave system. \textbf{Upper
panel}: Topological structures of each cluster, \textbf{Lower panel}:
\textbf{NR}-diagrams of the clusters shown in the upper panel.}
\end{figure*}
\begin{figure*}
\begin{center}
\includegraphics[width=12.0cm,height=2.0cm]{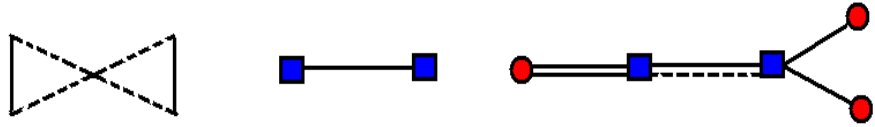}
\end{center}
\vskip -0.6cm
\caption{\label{f:4Graph} Examples of clusters in 4-wave systems. \textbf{Left}: Topological representation of a quartet. \textbf{Middle}: \textbf{NR}-diagram of a \textbf{V}-scale-scale cluster. \textbf{Right}:
\textbf{NR}-diagram of a mixed cascade cluster, with one \textbf{E}-angle-scale, one \textbf{VD}-scale-scale and two \textbf{V}-scale-angle connections.}
\end{figure*}

In the last decade, this standard view proved to be incomplete while finite-size effects are well observable within the inertial interval (\cite{DLN07,PZ00,ZKPD05} etc.). As it was shown in \cite{K06-1}, this is due to the notorious small divisor problem:
 each small divisor leaves a gap in the power spectrum, shown as an empty circle in Fig.\ref{f:LamTur}, lower panel. The radius $\mathcal{R}$ of each circle can be computed using the Thue-Siegel-Roth theorem \cite{K07} for a big class of dispersion functions, which covers in particularly various types of water waves, oceanic planetary waves, drift waves in laboratory plasma etc. Gaps corresponding to  exact and quasi-resonances are shown by yellow squares. Some gaps, shown by blue diamonds, correspond to  discrete modes that do not take part in resonances and just keep their initial energy on the corresponding time-scale \cite{K94}. Accordingly, 3 types of solutions of (\ref{3res}),(\ref{4res}) must be distinguished:

\textbf{I:} \emph{exact resonances},  $ \Omega =0;$

\textbf{II:} \emph{quasi-resonances},  $ 0< \mathcal{R} \le \Omega \wt{<} (\p \o /\p k) 2 \pi/ L;$

\textbf{III:} \emph{approximate interactions}, $\Omega \gg (\p \o
/\p k) 2 \pi/ L.$

\noindent  Types I and II are the subject of  DWT theory while Type III is covered by SWT theory.
Resonance clusters of the first two types include low-dimensional systems but \emph{are not reduced to} them: some clusters consist of hundreds or even thousands connected triads or quartets. These considerations justify the two-layer model of laminated turbulence, consisting of discrete and continuous (statistical) layers, presented in \cite{K06-1}, where the notion of DWT was first introduced.

\noindent {\bf 3. Exact resonances}.
3-wave and 4-wave resonance conditions have the form%
\vskip -0.6cm
 \bea %
\o_1+\o_2=\o_3+\Omega, \quad \mathbf{k}_1+\mathbf{k}_2=\mathbf{k}_3, \quad  \mbox{and} \label{3res}\\
\o_1+\o_2=\o_3+\o_4 + \Omega,\quad
\mathbf{k}_1+\mathbf{k}_2=\mathbf{k}_3+\mathbf{k}_4. \label{4res}
\eea%
\vskip -0.3cm
\noindent correspondingly. Minimal possible clusters in three- and four-wave systems are triads and quartets correspondingly, they are called \emph{primary clusters}. Cluster of arbitrary structure can be constructed from primary clusters, as well as its dynamical system. All constructions below for concreteness  are presented for \textbf{2D}-wave vectors.

\noindent {\bf 3.1. Resonance clustering in three-wave systems.}
To construct complete resonance clustering in a given spectral place,
three consequent steps have to be performed.

\emph{At the first step}, each resonant mode with
wave vector $\mathbf{k}=(m,n)$ is presented by a node of the two-dimensional integer
lattice $(m,n)$ and every three nodes constituting a solution of (\ref{3res}) are
connected by lines. The resulting graph is called
\emph{geometrical structure} (see \cite{KM07}, Fig.1, upper panel).

\emph{At the second step}, \emph{topological structure} is extracted, which consists
of all connected components found in the geometrical structure.
Primary clusters are shown as  triangles and their dynamical systems have standard form%
\vskip -0.6cm
\be \label{dyn3waves}  \dot{B}_1=   Z B_2^*B_3,\quad
\dot{B}_2= Z B_1^* B_3, \quad \dot{B}_3= -  Z B_1 B_2, %
\ee %%
\vskip -0.3cm
\noindent  differing only in magnitudes of the coupling coefficient $Z$. Due to  Hasselmann`s criterion of nonlinear wave instability \cite{Has67}, dynamical properties of the $\o_3$-mode with the highest frequency differ from those with smaller frequencies. We follow \cite{KL08} denoting $\o_3$-mode as \textbf{A}- and two other as \textbf{P}-modes.

\emph{At the third step}, \emph{a hyper-graph with marked arcs},  is constructed whose (hyper-)vertexes are triangles constructed at the previous step (see Fig.\ref{f:3Graph}, upper panel). To simplify graphical presentation for bigger clusters, we prefer \textbf{NR}-diagram presentation (NR for nonlinear resonance), without letters and arrows but with two types of half-edges instead: \emph{bold} for \textbf{A}-mode and \emph{dotted} - for \textbf{P}-mode.

Given a \textbf{NR}-diagram, the corresponding dynamical system is constructed by coupling 4, 3 or 2 systems of the form  (\ref{dyn3waves}) and equaling appropriate $B_i$ and $B_j$. For instance, the dynamical system of the two-triad cluster shown in Fig.\ref{f:3Graph} (it is unique, up to the change of indices $1 \leftrightarrow 2$ and $4 \leftrightarrow 5$) reads
\vskip -0.6cm
\bea \label{kite-1}
\begin{cases}
\dot{B}_1=   Z B_2^*B_3,\quad
\dot{B}_2= Z B_1^* B_3, \quad \dot{B}_3= -  Z B_1 B_2, \nonumber \\
  \dot{B}_4=   \widetilde{Z} B_5^*B_6,\quad
\dot{B}_5=  \widetilde{Z} B_4^* B_6, \quad \dot{B}_6= -   \widetilde{Z} B_4 B_5, \quad \RA \nonumber\\
B_1=B_4 \, (\mbox{\textbf{PP}-connection}), \, B_3=B_5 \, (\mbox{\textbf{AP}-connection})  \nonumber%
\end{cases}
\eea
\vskip -0.6cm
\bea \label{kite-2}
\begin{cases}
\dot{B}_1=   Z B_2^*B_3 + \widetilde{Z} B_3^*B_6,\quad
\dot{B}_2= Z B_1^* B_3, \nonumber \\
\dot{B}_3= -  Z B_1 B_2 + \widetilde{Z} B_1^* B_6,
\quad \dot{B}_6= -   \widetilde{Z} B_1 B_3,  \nonumber
\end{cases}
\eea
\vskip -0.3cm
We have studied  resonance clustering in  data sets computed for various 2D-wave systems (different types of water waves, oceanic and atmospheric planetary  waves, drift waves, etc.)
in the model spectral domain $|m|,\, |n| \le 10^3.$ In average, about $80-90\%$ of all clusters consist of only one or two connected triads, though bigger clusters of a few thousand connected triads were also observed. Some clusters have integrable dynamics \cite{BK,Ver}, depending on coupling coefficients $Z$ and/or on initial conditions and/or connection type, while dynamics of other clusters is chaotic.

\emph{A fact of the major importance} is: chaotic behavior is observable already in the systems consisting of two triads (\textbf{AA}-connection \emph{via} one vertex,  $Z/\widetilde{Z}=0.75$, work in progress). On the other hand, clusters of special form  may be integrable for arbitrary finite number of triads \cite{Ver}. This means that clusters with chaotic and integrable dynamics co-exist, in different sub-spaces of the $\mathbf{k}$-space.

Summarizing, an arbitrary cluster appearing in a 3-wave systems can be represented as a \textbf{NR}-diagram with 1 type of vertices and two types of marked edges: a single line for vertex-connection   and a double line for edge-connection. Each edge is marked as \textbf{AA}, \textbf{AP} or \textbf{PP}. The presentation is sufficient to reconstruct uniquely the dynamical system of an arbitrary cluster.

\noindent {\bf 3.2. Resonance clustering in four-wave systems.}
 Any 3-wave resonance generates energy transport over the scales in the  $\mathbf{k}$-space; this   is not the case for 4-wave systems where three types of energy transport are observed: a) over scales, b) over angles, and c) mixed cascades, including both scale- and angle-resonances \cite{K07}. Correspondingly, our \textbf{NR}-diagram representation of a 4-wave system has two types of vertices - circles for angle-resonances and squares for scale-resonances.  Another important dynamical difference between 3- and 4-wave systems is that for quartets, no generic criterion of instability is known \cite{Has67}; however, for the special case of   $\o_1=\o_2$ or  $\o_3=\o_4$, the 3-wave criterion can be used, of course. To keep information about the vertices corresponding to the terms on the left and right side of (\ref{4res}), in geometrical representation each quartet consists of two edges (bold lines) connecting vertices from the same side of (\ref{4res}) and
two diagonals (dashed lines). Accordingly, in \textbf{NR}-diagram representation,  connections are possible \emph{via}
 a vertex (a single bold line), edge (double bold), edge-diagonal (double, bold and dashed), diagonal-diagonal (double, dashed and dashed), notated as \textbf{V}-, \textbf{E}-, \textbf{ED}- and \textbf{DD}-connections correspondingly (see Fig.\ref{f:4Graph}). Notice that connection types \textbf{A} and \textbf{P} in a 3-wave system, shown in corresponding \textbf{NR}-diagrams, define \emph{dynamical characteristics}  of a connection. In a 4-wave system, connection types \textbf{V}, \textbf{E} and \textbf{D} do not hold dynamical information in the general case, only in some particular cases. 
 
\noindent {\bf 4. Diagrams for DWT \emph{versus} diagrams for SWT}. 
Various diagram techniques, beginning with Feynman diagrams, are widely used for description of STW. \textbf{NR}-diagram\index{diagram!\textbf{NR}-} 
is a tool to represent DWT; main  differences between these two types of diagrams can be formulated as follows: 
for a resonance cluster look somewhat similar to the Feynman diagrams\index{diagram!Feynman} known in quantum mechanics. The differences are as follows:
1) In a Feynman diagram each vertex represent a particle, which corresponds to \emph{a single wave} in the topological representation of a cluster, while in \textbf{NR}-diagram \emph{vertexes are primary clusters} -- resonant triads or quartets, depending on the order of resonance in the wave system under consideration.
2) Only \emph{a sum of all Feynman diagrams} represents possible interactions of a given particle with other particles. On the contrary, \emph{one \textbf{NR}-diagram} represents complete resonance clusters. 
3) A Feynman diagram \emph{does not allow computing the amplitudes} of the scattering process, it only gives a contribution corresponding to one term in the perturbation expansion. On the contrary, \textbf{NR}-diagram allows to reconstruct \emph{uniquely} dynamical system whose \emph{solutions are the amplitudes} of resonance cluster.

\noindent {\bf 5. Energy transport in DWT}.
The two-layer model of laminated turbulence gives only kinematic explanation for the generic co-existence of SWT and DWT. Numerical simulations with Hamiltonian dynamical equations \cite{ZKPD05} demonstrate the same. But we do not know anything about dynamical mechanisms underlying this co-existence. The problems to coop with are the following. In SWT, energy flow is  described by energy spectra $\mathbf{k}^{-a}$ which means, in particular that a) waves with wave-lengths of the same order also have energies of the same order; b) the magnitude of the initial energy of the system, $E_0$, is not important -- only the fact that $E_0=\const$ is used; and c) wave phases are random. \emph{Statements a)-c) do not hold in DWT.} Indeed, in many 3-wave systems waves with wave-lengths of order $k$ and $k^2$ can interact resonantly ("relaxed" locality, \cite{AMS98}), in 4-wave systems - waves with arbitrary big difference in wave-lengths can resonate (no locality, \cite{KK06}). This means that the magnitudes of modes' energies are not defined by the scales in THE $\mathbf{k}$-space but by the initial values of modes´ energies and phases. Importance of phases in low-dimensional systems has been realized long ago \cite{phases} but no analytical expression for phases was known.
The most compact expressions for energy and phase evolution in DWT regimes known to us presently read:
\be \label{EnDWT}
E_{mode}(T) \sim  \frac{1}{\mu}\Big[\sum_{i=0}^{\infty}\frac{q^{(2i+1)/2}}{1-q^{2i+1}}\sin(2i+1)\pi \, \frac{T-t_0}{\tau} \Big]^2,\ee
\bea \label{PhDWT}
\cot|\varphi(T)|\sim \frac{q}{Z^3\tau^3} \sum_{i=0}^{\infty}\frac{q^{i}}{1-q^{2i+1}}\sin(2i+1)\pi
\frac{T-t_0}{\tau}\nonumber \\
\times \sum_{j=0}^{\infty}\frac{(2j+1)\,q^{j}}{1-q^{2j+1}}\cos(2j+1)\pi \frac{T-t_0}{\tau}\,,
\eea
where $E_{mode}$ is THE energy of a resonant mode in a 3-wave system and $\varphi$ is the dynamical phase \cite{BK} which is a combination of initial phases corresponding to the resonance conditions (\ref{3res}). Here $Z$ is the coupling coefficient from (\ref{dyn3waves}) and depends only on wave-numbers; $\mu, \tau, t_0$ and $q$ are known expressions, including elliptic integrals, elliptic functions, their nomes and modula. They depend on wave-numbers, initial energy and initial energy distribution within a triad: not only $E_{0}$ is important but also the part of it contained in each mode. Expressions similar to (\ref{EnDWT}), (\ref{PhDWT}) can probably be obtained for a 4-wave system, along the lines given in \cite{SS05}. An important fact is that for a fixed resonant triad and fixed initial conditions, $Z,\mu, \tau, t_0, q $ \emph{are constants}, i.e. (\ref{EnDWT}), (\ref{PhDWT}) can be used directly for computing \emph{exactly} energy and phase evolution of  modes in primary clusters. These formulas can also be used for computing \emph{approximately} energy and phase evolution in  clusters of more general structure (for further discussion see \cite{K09}, Chapter 4).

\noindent {\bf 6. Summary}.\\
\textbullet~
O. Phillips pointed out in \cite{Ph81}  that SWT theory has reached its limitations and "new
physics, new mathematics and new intuition is required" for understanding discrete effects in wave turbulent systems. DWT theory gives reliable description of these effects and is indeed a natural counterpart to the classical SWT. In particular, this means that the classical notion of \emph{wave interactions} is not sufficient and might even be misleading for describing a generic wave turbulent system while it does not allow to distinguish between discrete and statistical regimes. Notions of \emph{exact resonances}, \emph{quasi-resonances} and \emph{approximate  interactions} should be used.

\textbullet~ DWT is characterized by the \emph{clustering} itself, and \emph{not by the number of modes} in particular clusters which can be fairly big. This means that though chaotic dynamics can be accounted for in DWT, it can not be generally described by kinetic equations: each cluster has its own independent dynamics, sometimes integrable, sometimes not. Clusters of a given form generate persisting patterns observable in laboratory experiments (\cite{HH}, gravity water waves) and in measured data (\cite{KL07}, atmospheric planetary waves). Their existence might give a better explanation of such well-known phenomena as Benjamin-Fair instability \cite{BF67} (for a profound discussion, see \cite{BF-dis}) or freak waves (work in progress).

\textbullet~ Numerical simulations \cite{Tan} show (for gravity water waves) that interactions of  Types I and II are observable on the linear time scale, not on the time scale $\mathcal{O}(t/\e^4)$ as predicted by SWT theory. This indicates that DWT theory might give more appropriate foundation for a short-term forecast.

\textbullet~ In some physical systems, only SWT is observable \cite{Grant} (on-cite measurements of tidal currents), in other - only DWT \cite{DLN07} (laboratory experiments). Co-existence of SWT and DWT has been demonstrated in numerical experiments \cite{ZKPD05} (gravity water waves) where this regime was called mesoscopic wave turbulence. Including  additional physical parameters can yield transition from SWT to DWT \cite{CK09} (capillary water waves, with and without rotation). Fluid mechanics examples do not exhaust, of course, the application areas of DWT, which include (similar to classical SWT) biology, medicine, astronomy, chemistry, sociology, etc.  The problem of the utmost importance presently is to gain more understanding about SWT $\rightleftarrows$ DWT energy transport.

{\bf Acknowledgements}. Author acknowledges the
supports of the Austrian Science Foundation (FWF) under project
P20164-N18.

\end{document}